\documentclass[12pt]{article}
\usepackage{amsmath,amssymb,epsfig}
\usepackage{graphicx}
\usepackage{epstopdf}
\usepackage{epsf}
\usepackage{makeidx}
\usepackage{cite}

\def\bPsi{\bar{\Psi}}

\def \cL{\mathcal{L}}
\def \CP3{\Bbb{C}\mathrm{P}^3}

\def \be{\begin{equation}}
\def \ee{\end{equation}}
\def \beq{\begin{eqnarray}}
\def \eeq{\end{eqnarray}}

\makeatletter

\renewcommand\section{\@startsection {section}{1}{\z@}%
                                   {-3.5ex \@plus -1ex \@minus -.2ex}%
                                   {2.3ex \@plus.2ex}%
                                   {\normalfont\large\bfseries}}

\renewcommand\subsection{\@startsection{subsection}{2}{\z@}%
                                     {-3.25ex\@plus -1ex \@minus -.2ex}%
                                     {1.5ex \@plus .2ex}%
                                     {\normalfont\normalsize\bfseries}}

\newcount\hour \newcount\minute
\hour=\time \divide \hour by 60
\minute=\time
\def\now{%
\ifnum \hour<13
  \ifnum \hour=0 \advance \hour by 12 \number\hour:\else \number\hour:\fi%
     \ifnum \minute<10 0\fi%
     \number\minute%
\ A.M.%
\else \advance \hour by -12 \number\hour:%
  \ifnum \minute<10 0\fi%
  \number\minute%
  \ P.M.%
\fi%
}

\makeatother

\begin{document}

\baselineskip=18pt  
\numberwithin{equation}{section}  
\allowdisplaybreaks  



%
%


\thispagestyle{empty}

\vspace*{-2cm}
\begin{flushright}
{\tt arXiv:0808.0500}\\
CALT-68-2696\\
IPMU-08-0052
\end{flushright}


\vspace*{1.7cm}
\begin{center}
 {\large {\bf Superconformal Chern-Simons Theories }\\ 
 {\bf and the Squashed Seven Sphere}}

 \vspace*{2cm}
 Hirosi Ooguri,$^{*,\dagger}$ and Chang-Soon Park,$^*$\\
 \vspace*{1.0cm}
 $^*$
{\it California Institute of Technology 452-48, Pasadena, CA 91125, USA}\\
~~\\
 $^\dagger$ {\it Institute for the Physics and Mathematics of the Universe, \\
University of Tokyo, Kashiwa, Chiba 277-8582, Japan}\\[1ex]
 \vspace*{0.8cm}


\end{center}
\vspace*{.5cm}

\noindent
We show that there are two supersymmetric completions of the three-dimensional Chern-Simons theory of level $k$ with gauge group
$U(N) \times U(N)$ coupled to four sets of massless scalars and spinors in the bi-fundamental representation, if we require $Sp(2) \subset SU(4)$ global symmetry with the matter fields in the fundamental representation of $SU(4)$. One is the $\mathcal{N}=6$ superconformal theory recently studied in
hep-th/0806.1218 and another is a new theory with $\mathcal{N}=1$ superconformal symmetry. We conjecture that the $\mathcal{N}=1$ theory is dual to M theory on
$AdS_4 \times {squashed}\, S^7/\Bbb{Z}_k$.

\newpage
\setcounter{page}{1} 





\section{Introduction}
In \cite{Schwarz:2004yj}, it was proposed that the low energy effective theories of coincident M2 branes are described by superconformal field theories in which Chern-Simons gauge fields couple to scalar and spinor fields.
Recently, the $\mathcal{N}=8$ superconformal Chern-Simons theory was discovered by Bagger and Lambert\cite{Bagger:2006sk, Bagger:2007jr, Bagger:2007vi}. A closely related work is \cite{Gustavsson:2007vu}. The theory has $SO(4)$ gauge symmetry. There have been numerous subsequent attempts to generalize the theory especially to extend to the gauge group other than $SO(4)$\cite{Bandres:2008vf,Papadopoulos:2008sk,Gauntlett:2008uf,Ho:2008ei,Bandres:2008kj,Gomis:2008be,Distler:2008mk,Benvenuti:2008bt}.
More recently, Aharony, Bergman, Jafferis and Maldacena\cite{Aharony:2008ug} discovered a family of $\mathcal{N}=6$ superconformal Chern-Simons theories with matter fields. Their construction contains the Bagger-Lambert theory as a special case.
In the $AdS/CFT$ context\cite{Maldacena:1997re,Witten:1998qj,Gubser:1998bc}, the theories are conjectured to be dual to M theory on $AdS_4 \times S^7/\Bbb{Z}_k$, and to type IIA string theory on $AdS_4 \times \CP3$ in the 't Hooft limit (large $N$ with $N/k$ fixed).
Soon after that, generalizations to various directions have been explored\cite{Benna:2008zy,Bhattacharya:2008bja,Nishioka:2008gz,Honma:2008jd,Imamura:2008nn,Minahan:2008hf,Armoni:2008kr,Gaiotto:2008cg,Ahn:2008gda,Grignani:2008is,Hosomichi:2008jb,Hanany:2008qc,Bagger:2008se,Terashima:2008sy}.

In this paper, we will start with the three-dimensional $U(N)\times U(N)$ Chern-Simons Lagrangian plus the kinetic terms for four boson and fermion matter fields in the bi-fundamental representation of the gauge group. Then we will supersymmetrize the Lagrangian in such a way that the resulting Lagrangian has $\mathcal{N}=1$ supersymmetry. In \cite{Bandres:2008ry}, it was shown that, if we require $SU(4)$ global R-symmetry, we end up with the $\mathcal{N}=6$ superconformal Chern-Simons theory constructed in \cite{Aharony:2008ug}.
Here, we will instead require that the Lagrangian has $\mathcal{N}=1$ supersymmetry with $Sp(2)\subset SU(4)$ global symmetry\footnote{Our notation is such that $Sp(1)$ has rank 1.  Thus $Sp(2) \cong SO(5) \subset SO(6) \cong SU(4)$. }. The supercharge is a singlet under $Sp(2)$. Then there is only one possible solution if we require that the Lagrangian carries no dimensionful parameters.
The moduli space is still $\Bbb{C}^4/\Bbb{Z}_k$ as in the $\mathcal{N}=6$ case, but the metric on the moduli space can be different. This theory is interesting since we know that, on the gravity side, there are precisely two solutions on $AdS_4\times S^7$\cite{Awada:1982pk,Duff:1983nu,Nilsson:1984bj}. One solution gives the usual round metric on $S^7$ and has $\mathcal{N}=8$ supersymmetry, which is broken to $\mathcal{N}=6$ after orbifolding by $\Bbb{Z}_k$. The isometry on the sphere reduces from $SO(8)$ to $SU(4)\times U(1)$. The other solution has the ``squashed" metric on $S^7$ and has $\mathcal{N}=1$ supersymmetry. The isometry on $S^7$ is $Sp(2)\times Sp(1)$. After orbifolding by $\Bbb{Z}_k$, we still have $\mathcal{N}=1$ supersymmetry, but the isometry is broken to $Sp(2)\times U(1)$. So we conjecture that the $\mathcal{N}=1$ superconformal Chern-Simons theory is dual to the supergravity solution with the squashed metric on the sphere.

In section 2, we introduce notation and show how we construct the $\mathcal{N}=1$ superconformal Chern-Simons theories with matter fields. In section 3, we review the supergravity solutions on $AdS_4\times S^7$ and their quotients, and relate them with the superconformal Chern-Simons-matter theories described in section 2.
In the Appendix, we explain the derivation of the $\mathcal{N}=1$ superconformal theories in more detail and show the invariance of the action under the superconformal transformation explicitly.

\section{Construction of the $\mathcal{N}=1$ superconformal Chern-Simons-matter theories}
In this section, we will construct the $\mathcal{N}=1$ superconformal Chern-Simons-matter theories in three dimensions with $Sp(2)\times U(1)$ global symmetry starting from the conformal field theories proposed in \cite{Aharony:2008ug}.

\subsection{Review of $\mathcal{N}=6$ superconformal Chern-Simons-matter theories}
First, let us present the $\mathcal{N}=6$ superconformal theory in three dimensions in the notation of \cite{Bandres:2008ry}.
The theory has the gauge group $U(N)\times U(N)$ and there are four complex scalars $(X_A)^a_{~\hat{a}}$ in the representation $(\overline{\bf{N}},\bf{N})$ under the gauge group and $(X^A)^{\hat{a}}_{~a}$ in $(\bf{N},\overline{\bf{N}})$ where $A=1,\cdots,4$. A lower index labels the $\bf{4}$ representation of the global $SU(4)$ R-symmetry and an upper index the complex-conjugate $\bar{\bf{4}}$. In the same way, we have the fermionic fields $(\Psi^A)^a_{~\hat{a}}$ and $(\Psi_A)^{\hat{a}}_{~a}$, which are two-component spinors. The bar on $\bar{\Psi}^A$ indicates transposition, followed by right multiplication by $\gamma^0$. Note that we do not take an additional complex conjugation. The $2\times 2$ Dirac matrices satisfy $\{\gamma^{\mu}, \gamma^{\nu} \}=2\eta^{\mu\nu}$ with $\eta^{\mu\nu}=\mathrm{diag}(-1,1,1)$. We will use a Majorana representation and choose a basis such that $\gamma^{\mu\nu\lambda}=\epsilon^{\mu\nu\lambda}$. For example, $\gamma^0=i\sigma^2$, $\gamma^1=\sigma^1$, and $\gamma^2=\sigma^3$. The $U(N)$ gauge fields are hermitian matrices $A^a_{~b}$ and $\hat{A}^{\hat{a}}_{~\hat{b}}$.
The covariant derivatives are
\begin{equation}
\begin{split}
D_{\mu} X_A &= \partial_{\mu} X_A + i(A_{\mu} X_A - X_A \hat{A}_{\mu})\\
D_{\mu} X^A &= \partial_{\mu} X^A + i(\hat{A}_{\mu} X^A - X^A A_{\mu})\;.
\end{split}
\end{equation}

The Lagrangian consists of several parts:
\begin{equation}\label{E:N6Lag}
\begin{split}
\cL_{kin}&=\frac{k}{2\pi} \mathrm{tr}\left(-D^{\mu} X^A D_{\mu} X_A + i\bar{\Psi}_A \gamma^{\mu} D_{\mu} \Psi^A\right)\\
\cL_{CS}&=\frac{k}{2\pi} \epsilon^{\mu\nu\lambda} \mathrm{tr} \left(\frac 1 2 A_{\mu} \partial_{\nu} A_{\lambda} + \frac i 3 A_{\mu}A_{\nu}A_{\lambda} - \frac 1 2 \hat{A}_{\mu} \partial_{\nu} \hat{A}_{\lambda} - \frac i 3 \hat{A}_{\mu}\hat{A}_{\nu}\hat{A}_{\lambda}\right)\\
\cL_{4a}&=\frac{k}{2\pi} \left[i\epsilon^{ABCD} \mathrm{tr}\left(\bar{\Psi}_A X_B \Psi_C X_D\right)-i\epsilon_{ABCD} \mathrm{tr}\left(\bar{\Psi}^A X^B \Psi^C X^D\right)\right]\\
\cL_{4b}&=\frac{k}{2\pi} i \mathrm{tr} \left(\bar{\Psi}^A \Psi_A X_B X^B - \bar{\Psi}_A \Psi^A X^B X_B\right)\\
\cL_{4c}&=\frac{k}{2\pi} 2i \mathrm{tr}\left(\bar{\Psi}_A \Psi^B X^A X_B - \bar{\Psi}^B \Psi_A X_B X^A \right)\\
\cL_{pot}&=\frac{k}{2\pi} \frac 1 3 \mathrm{tr} \Big[ X^A X_A X^B X_B X^C X_C + X_A X^A X_B X^B X_C X^C \\
            &\qquad\qquad +4X_A X^B X_C X^A X_B X^C -6 X^A X_B X^B X_A X^C X_C \Big ]\;.
\end{split}
\end{equation}
Note that we assume $k$ is positive to give the correct sign for the $X$ field kinetic term. When $k$ is negative, the signs of the first two terms in $\cL_{kin}$ will change and the other terms change appropriately, in addition to suitable changes in the supersymmetry transformation rules.

The supersymmetry transformation is given by
\begin{equation}\label{E:SUSYTrans}
\begin{split}
\delta X_A &= i \Gamma^I_{AB}\, \bar{\epsilon}^I\, \Psi^B\\
\delta X^A &= -i \tilde{\Gamma}^{IAB}\bar{\Psi}_B \,\epsilon^I\\
\delta \Psi_A &= \Gamma^I_{AB} \gamma^{\mu} \,\epsilon^I\, D_{\mu} X^B + \delta_3 \Psi_A\\
\delta \Psi^A &= -\tilde{\Gamma}^{IAB} \gamma^{\mu}\, \epsilon^I \, D_{\mu} X_B + \delta_3 \Psi^A\\
\delta A_{\mu} &= \Gamma^I_{AB} \,\bar{\epsilon}^I \,\gamma_{\mu} \Psi^{A} X^B - \tilde{\Gamma}^{IAB} X_B \bar{\Psi}_A \gamma_{\mu}\, \epsilon^I\\
\delta \hat{A}_{\mu} &= \Gamma^I_{AB} X^B \,\bar{\epsilon}^I\, \gamma_{\mu} \Psi^A - \tilde{\Gamma}^{IAB} \bar{\Psi}_A \gamma_{\mu}\,\epsilon^I\, X_B\;,
\end{split}
\end{equation}
where
\begin{equation}\label{E:delta3}
\begin{split}
\delta_3 \Psi^A &= [\tilde{\Gamma}^{IAB} (X_C X^C X_B - X_B X^C X_C) - 2\tilde{\Gamma}^{IBC} X_B X^A X_C ]\epsilon^I\\
\delta_3 \Psi_A &= [\Gamma^I_{AB}(X^C X_C X^B - X^B X_C X^C)-2\Gamma^I_{BC} X^B X_A X^C]\epsilon^I\;.
\end{split}
\end{equation}
Here $I$ runs from 1 to 6 and labels the $\bf{6}$ representation of $SO(6)$. $\Gamma^I_{AB}$ is the Clebsch-Gordan coefficient that transforms two $\bf{4}$s into $\bf{6}$.
$\Gamma^{I}_{AB}=-\Gamma^{I}_{BA}$ and $\tilde{\Gamma}^I=(\Gamma^I)^{\dagger}$.
Note that there is a global $U(1)$ symmetry under which $X_A$ and $\Psi^A$ has charge +1 and $X^A$ and $\Psi_A$ charge -1. The total global symmetry is $SU(4)_R \times U(1)$.
Let us briefly mention how the supersymmetries of \eqref{E:N6Lag} are preserved\cite{Bandres:2008ry}. Supersymmetric variations of $\cL_{kin}$ and $\cL_{CS}$ almost cancel out. But there are some remaining terms that require the additional terms in the Lagrangian. The variation due to $\delta A_{\mu}$ in the spinor kinetic term in $\cL_{kin}$ is canceled out by varying the $X$ fields in $\cL_4=\cL_{4a}+\cL_{4b}+\cL_{4c}$. The variations due to $\delta A_{\mu}$ in the $X$ kinetic term in $\cL_{kin}$ and $\delta \Psi$(without $\delta_3 \Psi$) in $\cL_4$ are canceled out by the variation of $\delta_3 \Psi$ in the spinor kinetic term if we choose the variation $\delta_3 \Psi^A$ and $\delta_3 \Psi_A$ as shown in \eqref{E:delta3}. The $\delta_3 \Psi$ variation of $\cL_4$ is canceled out by the $X$ variation of $\cL_{pot}$. So the whole Lagrangian is supersymmetric.

\subsection{$\mathcal{N}=1$ superconformal Chern-Simons-matter theories}
Here, we will construct $\mathcal{N}=1$ superconformal field theory with $Sp(2)\times U(1)$ symmetry.
First, let us impose the $Sp(2)$ invariance condition.
Note that $Sp(2)$ is the intersection of $SU(4)$ and $Sp(4,\Bbb{C})$. Therefore, we have an invariant antisymmetric $4\times 4$ tensor $\Omega_{AB}$ under $Sp(2)$.
Also, we expect the supersymmetry is reduced from $\mathcal{N}=6$ to $\mathcal{N}=1$. Since $\Gamma^I_{AB}$ for each $I$ is a non-degenerate antisymmetric $4\times 4$ tensor, a natural way to proceed is to look for a theory in which the supersymmetric transformation is given by \eqref{E:SUSYTrans} with $\Gamma^I_{AB} \, \epsilon^I$ replaced by  $\epsilon \, \Omega_{AB}$ with a spinor $\epsilon$ that is a singlet under $Sp(2)\cong SO(5)$. We will also define $\Omega^{AB}$ such that $\Omega^{AB} \Omega_{AC}=\delta^B_C$. For example, we can use $\Gamma^1=i\sigma_2 \otimes 1$, $\epsilon^I=\epsilon(1,0,0,0,0,0)$, $\Omega^{AB}=\Omega_{AB}=\Gamma^1$.

Since we have an additional antisymmetric invariant tensor $\Omega_{AB}$ compared to $SU(4)$ symmetric case, additional terms are allowed in the Lagrangian. For example, terms such as $\Omega^{AD} \Omega_{BC} \bar{\Psi}_A \Psi^B X^C X_D$ are allowed.
The most general possible forms with no dimensionful parameters are found in the appendix.
Starting from $\cL_{kin}+\cL_{CS}$ in \eqref{E:N6Lag}, we can look for a suitable linear combination of $\cL_{a,b,c}$ and $\cL_{pot}$ together with $\cL'$ that is invariant under supersymmetry. There are only two possible solutions. One is the $\mathcal{N}=6$ superconformal Chern-Simons theory with matter fields constructed in \cite{Aharony:2008ug} in the notation of \cite{Bandres:2008ry}. The other is the $\mathcal{N}=1$ theory whose Lagrangian is given by
\begin{equation}\label{E:N1Lag}
\begin{split}
\cL&=\frac{k}{2\pi} \mathrm{tr} \Big [ -D^{\mu} X^A D_{\mu} X_A + i\bar{\Psi}_A \gamma^{\mu} D_{\mu} \Psi^A \\
    &+ \epsilon^{\mu\nu\lambda} \left(\frac 1 2 A_{\mu} \partial_{\nu} A_{\lambda} + \frac i 3 A_{\mu}A_{\nu}A_{\lambda} - \frac 1 2 \hat{A}_{\mu} \partial_{\nu} \hat{A}_{\lambda} - \frac i 3 \hat{A}_{\mu}\hat{A}_{\nu}\hat{A}_{\lambda}\right)\\
    &-i\bPsi^A \Psi_A X_B X^B + i \bPsi_A \Psi^A X^B X_B\\
    &-2i\Omega^{AD} \Omega_{BC} \bPsi_A \Psi^B X^C X_D + 2i \Omega_{AD} \Omega^{BC} \bPsi^A \Psi_B X_C X^D\\
    &- X_A X^A X_B X^B X_C X^C - X^A X_A X^B X_B X^C X_C + 2 X^A X_B X^B X_A X^C X_C \Big ]\;.
\end{split}
\end{equation}
In the appendix, it is shown explicitly that classically the action is invariant under superconformal symmetry as well as supersymmetry.
Due to the presence of the antisymmetric tensor $\Omega_{AB}$ in the Lagrangian \eqref{E:N1Lag}, it is clear that no other supersymmetries will be preserved.
Note that the bosonic potential can be written in the form
\begin{equation}
V=\frac{k}{2\pi} \mathrm{tr}\left(N^A N_A\right)\;,
\end{equation}
where
\begin{equation}\label{E:NA}
\begin{split}
N^A &= \Omega^{AB} (X_C X^C X_B - X_B X^C X_C)\\
N_A &= \Omega_{AB} (X^B X_C X^C - X^C X_C X^B)
\end{split}
\end{equation}
are factors that appear in the supersymmetric transformation $\delta_3 \Psi^A$ and $\delta_3 \Psi_A$ as shown in \eqref{E:N1spintrans}.
Therefore the bosonic potential is manifestly positive definite and the classical moduli space is given by the solution $N^A=N_A=0$.
This condition is satisfied when $X$ fields are diagonalized.
It is straightforward to check that all off-diagonal excitations are generically massive.
Therefore the gauge symmetry is generically broken to $U(1)^N\times U(1)^N$ up to permutations of the diagonal elements.
Note that the moduli space is supersymmetric since $N_A=0$ implies $\delta_3 \Psi_A = 0$, which in turn implies $\delta \Psi_A=0$ in the vacuum.
For each $U(1)\times U(1)$, the matter fields are charged under only one linear combination of the $U(1)$'s.
But the $U(1)$ that couples to the matter fields do not preserve the gauge symmetry in the presence of the Chern-Simons terms, and instead the gauge symmetry reduces to $\Bbb{Z}_k$ due to flux quantization conditions \cite{Aharony:2008ug}.
Hence the classical moduli space is given by $(\Bbb{C}^4 / \Bbb{Z}_k)^N$ up to permutations. But the Lagrangian has only $Sp(2)\times U(1)$ symmetry due to the terms with the antisymmetric tensor $\Omega_{AB}$. Therefore, although the classical moduli space does not see any $Sp(2)\times U(1)$ structure, the low energy effective theory will have a non-trivial metric on the target space with $Sp(2) \times U(1)$ symmetry.

One may worry that the conformal invariance of the classical action may be broken by quantum effects. It turns out that there is no marginal operator besides the Lagrangian itself and that the only relevant operators consistent with supersymmetry are the mass terms in the combination
\begin{equation}\label{E:massterms}
\mathrm{tr} [i m \bPsi_A \Psi^A -m^2 X^A X_A - m(X_A X^A X_B X^B - X^A X_B X^B X_B) ]\;.
\end{equation}
To make this combination supersymmetric, one needs to modify the supersymmetry transformation by adding
$\delta' \Psi^A = -m \Omega^{AB} \epsilon X_B$
to the fermion transformation.
However, such terms cannot be generated perturbatively if one assumes that supersymmetry is unbroken. The flat directions parameterized by diagonal $X$'s represent supersymmetric vacua, and the standard argument shows that they are not lifted by perturbative effects. On the other hand, the mass terms would lift these vacua, except for the one at the origin. Thus, no relevant operators are generated perturbatively. We also note that the level $k$ is not shifted at one-loop since the field content of the $\mathcal{N}=1$ theories is the same as that of the $\mathcal{N}=6$ theories, where $k$ is not shifted \cite{Aharony:2008ug}.

It is also interesting to check whether a similar construction can yield a $\mathcal{N}=5$ supersymmetric Lagrangian for which the supercharges are in the $\bf 5$ of $SO(5)\cong Sp(2)$ representation. It turns out that we are not able to construct a solution. The procedure is the same as the previous situation and a sketchy description of the calculation is in the appendix. This may be related to the fact that there does not exist a supergravity solution on $AdS_4 \times S^7/\Bbb{Z}_k$ with $\mathcal{N}=5$ supersymmetry\cite{Awada:1982pk}.

\section{Dual M-theory Description}
Suppose the eleven dimensional spacetime is given in the form $\Bbb{R}^3 \times X$ where $X$ is an eight-dimensional cone over $S^7/\Bbb{Z}_k$, but with the squashed metric on it. The $\mathcal{N}=1$ superconformal theory can be obtained by placing $N$ M2-branes on the tip of the cone \cite{Klebanov:1999tb,Morrison:1998cs,Acharya:1998db,Duff:1995wk}, which is a singular $Spin(7)$ manifold\footnote{M theory on a class of $Spin(7)$ manifolds was studied in \cite{Gukov:2001hf,Gukov:2002zg}.}.
We propose that these superconformal theories are the $\mathcal{N}=1$ Chern-Simons-matter theories constructed in the previous section.
Note that the cone over $S^7$ has $\mathcal{N}=1$ supersymmetry, whose supercharge is a singlet under the isometry $Sp(2)\times Sp(1)$ of the squashed $S^7$. The orbifolding does not project out this singlet since $\Bbb{Z}_k$ acts on the $U(1)$ subgroup of $Sp(1)$.

The near horizon geometry of these M2 branes is $AdS_4 \times S^7/\Bbb{Z}_k$ with the squashed metric on $S^7/\Bbb{Z}_k$. The isometry of the squashed $S^7$ is $Sp(2)\times Sp(1)$, which is broken to $Sp(2)\times U(1)$ by $\Bbb{Z}_k$. We note that it is identical to the global symmetry of the $\mathcal{N}=1$ superconformal Chern-Simons theories.

The supergravity solution on $AdS_4 \times S^7$ with the squashed metric on $S^7$ is given by\cite{Awada:1982pk}
\begin{equation}\label{E:SUGRASol}
\begin{split}
ds^2 &= \frac{R^2} 4 ds^2_{AdS_4} + R^2 ds^2_{S^7}\\
F_4 &\sim N' \epsilon_4\\
R&=(2^5 \pi^2 N')^{1/6} l_p\;,
\end{split}
\end{equation}
where the metrics $ds^2_{AdS_4}$ and $ds^2_{S^7}$ have unit radius. One way to specify the metric on $S^7$ is to use the Fubini-Study metric on $P_2(H)$, the quaternionic projective plane. We choose a level surface of distance $r$ from a point in $P_2(H)$. This distance $r$ determines the degree of distortion: near $r=0$, the metric is almost round and it gets distorted as $r$ becomes large. The induced metric of the Fubini-Study metric on this seven dimensional surface defines the squashed metric. Explicitly,
\begin{equation}
ds^2_{S^7} = \kappa^2 (d\mu^2 + \frac 1 4 \omega_i ^2 \sin^2\mu  + \frac 1 4 \lambda^2 (\nu_i + \omega_i  \cos\mu )^2)\;,
\end{equation}
where $\kappa$ is the overall constant to be chosen later, and $\lambda$ is related to the distance $r$ such that $\lambda^2=\frac 1 {1+r^2}$, which parameterizes the degree of distortion.
The one-forms $\nu_i$ and $\omega_i$, $i=1,2,3$, are defined by
\begin{equation}
\nu_i = \sigma_i + \Sigma_i , \quad \omega_i = \sigma_i - \Sigma_i\;,
\end{equation}
with $\sigma_i$ and $\Sigma_i$ satisfying
\begin{equation}\label{E:leftsquash}
d\sigma_1=-\sigma_2 \wedge \sigma_3\;,\quad d\Sigma_1=-\Sigma_2 \wedge \Sigma_3\;,
\end{equation}
etc.
When $\lambda^2=1$, the metric is that of the round sphere, which has $SO(8)$ isometry. For all other $\lambda^2$, the isometry is $Sp(2)\times Sp(1)$. It is not generally an Einstein metric but it becomes so when $\lambda^2=1$ or $1/5$. When $\lambda^2=1/5$, there is only one Killing spinor, so it has $\mathcal{N}=1$ supersymmetry. It has the weak $G_2$ holonomy.
The overall constant $\kappa$ is chosen to satisfy $R_{S^7 \, mn} = 6 \delta_{mn}$: $\kappa^2=\frac 1 4$ for $\lambda^2=1$ and $\kappa^2= \frac 9 {20}$ for $\lambda^2=\frac 1 5$.
The two supergravity solutions are classically stable under the changes of the size and squashing parameters of $S^7$ \cite{Page:1984qv}.
There is actually a static domain wall interpolating the two solutions\cite{Ahn:1999dq}.

Since we want to quotient $S^7$ by $\Bbb{Z}_k$, it is more convenient to write the metric in a form that shows that $S^7$ is an $S^1$ bundle over $\CP3$. Then the metric has the form\cite{Nilsson:1984bj}
\begin{equation}\label{E:fibermetric}
ds^2_{S^7} = (d\phi' + \omega)^2 + ds^2_{\CP3}\;,
\end{equation}
where $\omega$ is a potential for a non-trivial topology on $\CP3$ and $\phi'$ is the periodic coordinate with period $2\pi$. $\CP3$ also admits a family of homogeneous metric labeled by $\lambda$\cite{Ziller}, for which the U(1) fibration over $\CP3$ gives the squashed $S^7$ with the same parameter $\lambda$. $\lambda^2=1$ is the standard Fubini-Study Einstein metric on $\CP3$ and gives the round seven-sphere metric when put in \eqref{E:fibermetric}. For other choices of $\lambda^2$, the corresponding metric is non-Einstein except at $\lambda^2=1/2$.
The supergravity solution on the squashed $S^7$ corresponds to \eqref{E:fibermetric} with $\lambda^2=1/5$. Interestingly the metric on $\CP3$ is not Einstein.

Given the form \eqref{E:fibermetric}, it is easy to take the $\Bbb{Z}_k$ quotient\cite{Aharony:2008ug}. We set $\phi'=\phi/k$ with $\phi=\phi+2\pi$. Then the metric is
\begin{equation}
ds^2_{S^7/{\Bbb{Z}_k}} =\frac 1 {k^2} (d\phi + k \omega)^2 + ds^2_{\CP3}\;.
\end{equation}
Since the volume of $S^7$ is reduced by a factor of $k$, the supergravity solution on $AdS_4\times S^7$ is obtained by setting $N'=k N$ and replacing $ds^2_{S^7}$ by $ds^2_{S^7/{\Bbb{Z}_k}}$ in \eqref{E:SUGRASol}. The supersymmetry is still $\mathcal{N}=1$ since the Killing spinor is a singlet under $Sp(2)\times Sp(1)$.

Let us mention that, when $k$ becomes its negative, $N'$ goes to $-N'$ and both $\mathcal{N}=6$ round-sphere and $\mathcal{N}=1$ squashed-sphere supergravity solutions reduce to $\mathcal{N}=0$ \cite{Awada:1982pk}. The supersymmetry becomes again $\mathcal{N}=6$ or $\mathcal{N}=1$ if we exchange the ${\bf 8}_s$ and ${\bf 8}_c$ representations of $SO(8)$, of which $SU(4)\times U(1)$ and $Sp(2)\times Sp(1)$ are subgroups, and change the left-squashed sphere to the right-squashed one in $\mathcal{N}=1$ case, which flips the minus signs in \eqref{E:leftsquash}. Let us see what this corresponds to in the field theory side.
Note that the sign of the bosonic kinetic term of \eqref{E:N6Lag} or \eqref{E:N1Lag} changes when $k$ becomes its negative so that the kinetic and Chern-Simons terms in the Lagrangian become
\begin{equation}
\begin{split}
&\frac{k}{2\pi} \mathrm{tr} \Big [ D^{\mu} X^A D_{\mu} X_A - i\bar{\Psi}_A \gamma^{\mu} D_{\mu} \Psi^A \\
    &+ \epsilon^{\mu\nu\lambda} \left(\frac 1 2 A_{\mu} \partial_{\nu} A_{\lambda} + \frac i 3 A_{\mu}A_{\nu}A_{\lambda} - \frac 1 2 \hat{A}_{\mu} \partial_{\nu} \hat{A}_{\lambda} - \frac i 3 \hat{A}_{\mu}\hat{A}_{\nu}\hat{A}_{\lambda}\right) \Big ] \;.
\end{split}
\end{equation}
The relative sign between boson and fermion matter fields is determined by supersymmetry. All the remaining terms change up to appropriate signs.
In this form, the original supersymmetry transformation in each case ceases to be a symmetry of the Lagrangian. Instead, a different supersymmetry such that
\begin{equation}
\delta A_{\mu}\;, \delta \hat{A}_{\mu}\;, \delta_3 \Psi^A \rightarrow - \delta A_{\mu}\;, - \delta \hat{A}_{\mu} \;,-\delta_3 \Psi^A
\end{equation}
becomes a symmetry of the Lagrangian.

\section{Conclusions}
In this paper, we started with three-dimensional $U(N)\times U(N)$ Chern-Simons theories with bi-fundamental bosonic and fermionic matter fields in $\bf{4}$ and $\bar{\bf{4}}$ of $SU(4)$. We then supersymmetrize this Lagrangian. If the final Lagrangian is to be invariant under $\mathcal{N}=6$ supersymmetry with $SU(4)_R\times U(1)$ global symmetry, we end up with the Lagrangian in \cite{Aharony:2008ug,Bandres:2008ry}. If we loosen the condition so that the final Lagrangian has $\mathcal{N}=1$ supersymmetry with $Sp(2)\times U(1) \subset SU(4)\times U(1)$ global symmetry, we have the Lagrangian \eqref{E:N1Lag} in addition to the previous $\mathcal{N}=6$ Lagrangian. Both have the same classical moduli space. The situation is very similar to the supergravity side since there are also two possible solutions on $AdS_4\times S^7/\Bbb{Z}_k$. In one case, the metric on the sphere is the usual round one, whereas in the other case, we have the squashed sphere. Therefore we propose that the $\mathcal{N}=1$ superconformal Chern-Simons-matter theory with the Lagrangian \eqref{E:N1Lag} describes $N$ M2-branes on the tip of the cone with squashed $S^7/\Bbb{Z}_k$ base in M-theory.

\section{Acknowledgments}
We would like to thank J. Schwarz for discussions.
C.P. thanks the students and organizers of Prospects in Theoretical Physics 2008 where the work was initiated.
H.O. thanks the Aspen Center for Physics for the hospitality.

This work is supported in part by DOE grant DE-FG03-92-ER40701. The work of H.O. is
also supported in part by a Grant-in-Aid for Scientific Research (C) 20540256 from the Japan
Society for the Promotion of Science, by the World Premier International Research Center
Initiative of MEXT of Japan, and by the Kavli Foundation. C.P. is supported in part by Samsung Scholarship.

\appendix
\section{Detailed construction of the $\mathcal{N}=1$ superconformal theories}
In this appendix, we show how we arrive at the $\mathcal{N}=1$ superconformal Chern-Simons Lagrangian \eqref{E:N1Lag} starting from the Chern-Simons term and the bosonic and fermionic matter terms $\cL_{kin}+\cL_{CS}$ in \eqref{E:N6Lag}. Since we have only $Sp(2)\subset SU(4)$ symmetry in the Lagrangian, there are additional terms allowed in the Lagrangian. The most general $\Psi\Psi X X$ combination of marginal operators is
\begin{equation}\label{E:sp2inv}
\begin{split}
\cL'=&a_1 \Omega^{AD} \Omega_{BC} \bar{\Psi}_A \Psi^B X^C X_D + a_2 \Omega_{AD}\Omega^{BC} \bPsi^A \Psi_B X_C X^D\\
    &+a_3 \Omega^{AC} \Omega^{BC} \bPsi_A X_B \Psi_C X_D + \bar{a}_3 \Omega_{AC} \Omega_{BD} \bPsi^A X^B \Psi^C X^D\\
    &+a_4 \Omega^{AB} \Omega^{CD} \bPsi_A X_B \Psi_C X_D + \bar{a}_4 \Omega_{AB} \Omega_{CD} \bPsi^A X^B \Psi^C X^D\;.
\end{split}
\end{equation}
There is also a part of the Lagrangian which consists of 6 $X$ fields such as $\Omega\Omega X X X X X X$, which we call $\cL''$.

We will deform \eqref{E:N6Lag} by varying coefficient for each term in $\cL_{4a,b,c}$ and $\cL_{pot}$.
So the Lagrangian we consider is the sum of
\begin{equation}
\begin{split}
\cL_{kin}&=\frac{k}{2\pi} \mathrm{tr}\left(-D^{\mu} X^A D_{\mu} X_A + i\bar{\Psi}_A \gamma^{\mu} D_{\mu} \Psi^A\right)\\
\cL_{CS}&=\frac{k}{2\pi} \epsilon^{\mu\nu\lambda} \mathrm{tr} \left(\frac 1 2 A_{\mu} \partial_{\nu} A_{\lambda} + \frac i 3 A_{\mu}A_{\nu}A_{\lambda} - \frac 1 2 \hat{A}_{\mu} \partial_{\nu} \hat{A}_{\lambda} - \frac i 3 \hat{A}_{\mu}\hat{A}_{\nu}\hat{A}_{\lambda}\right)\\
\cL_{4a}&=\frac{k}{2\pi} \left[i\bar \alpha_1 \epsilon^{ABCD} \mathrm{tr}\left(\bar{\Psi}_A X_B \Psi_C X_D\right)-i \alpha_1 \epsilon_{ABCD} \mathrm{tr}\left(\bar{\Psi}^A X^B \Psi^C X^D\right)\right]\\
\cL_{4b}&=\frac{k}{2\pi} i \mathrm{tr} \left(\alpha_{2,1} \bar{\Psi}^A \Psi_A X_B X^B - \alpha_{2,2} \bar{\Psi}_A \Psi^A X^B X_B\right)\\
\cL_{4c}&=\frac{k}{2\pi}  2i \mathrm{tr}\left(\alpha_{3,1} \bar{\Psi}_A \Psi^B X^A X_B - \alpha_{3,2} \bar{\Psi}^B \Psi_A X_B X^A \right)\\
\cL_{pot}&=\frac{k}{2\pi} \frac 1 3 \mathrm{tr} \Big[ \alpha_{4,1} X^A X_A X^B X_B X^C X_C + \alpha_{4,2} X_A X^A X_B X^B X_C X^C \\
            &\qquad\qquad +4 \alpha_{4,3} X_A X^B X_C X^A X_B X^C -6 \alpha_{4,4} X^A X_B X^B X_A X^C X_C \Big ]\;.
\end{split}
\end{equation}
with the addition of $\cL'$ and $\cL''$.

We now check under what condition the Lagrangian satisfies $\mathcal{N}=1$ supersymmetry given by
\begin{equation}\label{E:SUSYTrans1}
\begin{split}
\delta X_A &= i \Omega_{AB}\, \bar{\epsilon}\, \Psi^B\\
\delta X^A &= i \Omega^{AB}\bar{\Psi}_B\, \epsilon\\
\delta \Psi_A &= \Omega_{AB} \gamma^{\mu}\, \epsilon\, D_{\mu} X^B + \delta_3 \Psi_A\\
\delta \Psi^A &= \Omega^{AB} \gamma^{\mu}\, \epsilon\, D_{\mu} X_B + \delta_3\Psi^A\\
\delta A_{\mu} &= \Omega_{AB}\, \bar{\epsilon}\, \gamma_{\mu} \Psi^{A} X^B + \Omega^{AB} X_B \bar{\Psi}_A \gamma_{\mu} \, \epsilon\\
\delta \hat{A}_{\mu} &= \Omega_{AB} X^B \, \bar{\epsilon}\, \gamma_{\mu} \Psi^A +\Omega^{AB} \bar{\Psi}_A \gamma_{\mu}\, \epsilon X_B\;,
\end{split}
\end{equation}
where $\delta_3$ variation is to be determined.

Let's first vary $A$ field in the spinor kinetic term in $\cL_{kin}$. This yields a term
\begin{equation}
\frac k {2\pi} 2\,\Omega_{BC} \mathrm{tr} \left[ \, \bar{\epsilon} \, \Psi^A (\bPsi_A \Psi^B X^C - X^C \bPsi^B \Psi_A) \right]\;.
\end{equation}
The same term is generated by varying $X^B$ in the second term in $\cL_{4a}$. Such terms can arise in the terms with $a_1$ and $a_2$ coefficients in $\cL'$ by varying $X_B$ with the constraint that $a_2=-a_1$. Then all such terms cancel out when
\begin{equation}
2-2\alpha_1 + i a_2 =0\;.
\end{equation}
Then the variation $\delta_A (\Psi D \Psi) + \delta_A \cL_{4a}+\delta_{X} \cL'$ vanishes when
\begin{equation}\label{E:CoefRel}
\begin{split}
&\alpha_{2,1}=\alpha_{2,2}\equiv \alpha_2,\quad \alpha_{3,1}=\alpha_{3,2}\equiv \alpha_3\\
&\alpha_2=2\alpha_1-1,\quad \alpha_3=\alpha_1,\quad i a_2 = 2\alpha_1 - 2\;,
\end{split}
\end{equation}
and $a_3=a_4=0$.

Next, we consider the $\delta A_{\mu}$ variation in the $X$ field kinetic term and $\delta \Psi$(without $\delta_3 \Psi$) in $\cL_{4a,b,c}$ and $\cL'$.
Thess variations cancel against the $\delta_3 \Psi$ variation in the spinor kinetic term if we choose
\begin{equation}
\begin{split}
\delta_3 \Psi^A &= -\Omega^{AB} \,\epsilon\, (2\alpha_1 -1)(X_C X^C X_B - X_B X^C X_C ) + 2\alpha_1 \Omega^{BC} \,\epsilon\, X_B X^A X_C\\
\delta_3 \Psi_A &= -\Omega_{AC} \,\bar{\epsilon}\, (2\alpha_1-1)(X^C X_D X^D - X^D X_D X^C) + 2 \alpha_1 \,\bar{\epsilon}\, \Omega_{HK} X^K X_A X^H\;.
\end{split}
\end{equation}
Then the variations in $\cL_{4a,b,c}$ due to $\delta_3 \Psi$ have the form of the variation of terms with six $X$ fields, plus some additional terms, which vanish when $\alpha_1(\alpha_1-1)=0$. That is, when $\alpha_1=0$ or $\alpha_1=1$. The case with $\alpha_1=1$ is the $\mathcal{N}=6$ superconformal field theory. When $\alpha_1=0$, remembering \eqref{E:CoefRel}, we have the Lagrangian \eqref{E:N1Lag}:
\[
\begin{split}
\cL&=\frac{k}{2\pi} \mathrm{tr} \Big [ -D^{\mu} X^A D_{\mu} X_A + i\bar{\Psi}_A \gamma^{\mu} D_{\mu} \Psi^A \\
    &+ \epsilon^{\mu\nu\lambda} \left(\frac 1 2 A_{\mu} \partial_{\nu} A_{\lambda} + \frac i 3 A_{\mu}A_{\nu}A_{\lambda} - \frac 1 2 \hat{A}_{\mu} \partial_{\nu} \hat{A}_{\lambda} - \frac i 3 \hat{A}_{\mu}\hat{A}_{\nu}\hat{A}_{\lambda}\right)\\
    &-i\bPsi^A \Psi_A X_B X^B + i \bPsi_A \Psi^A X^B X_B\\
    &-2i\Omega^{AD} \Omega_{BC} \bPsi_A \Psi^B X^C X_D + 2i \Omega_{AD} \Omega^{BC} \bPsi^A \Psi_B X_C X^D\\
    &- X_A X^A X_B X^B X_C X^C - X^A X_A X^B X_B X^C X_C + 2 X^A X_B X^B X_A X^C X_C \Big ]\;.
\end{split}
\]
The $\delta_3 \Psi_A$ and $\delta_3 \Psi^A$ in the supersymmetry transformation for the spinors are then given by
\begin{equation}\label{E:N1spintrans}
\delta_3 \Psi^A = N^A \, \epsilon \;, \quad \delta_3 \Psi_A = N_A \, \epsilon \;,
\end{equation}
where $N^A$ and $N_A$ are defined in \eqref{E:NA}:
\[
\begin{split}
N^A &= \Omega^{AB} (X_C X^C X_B - X_B X^C X_C )\\
N_A &= \Omega_{AB} (X^B X_D X^D - X^D X_D X^B)\;.
\end{split}
\]

Let us show that the theory has the superconformal symmetry. Following the expressions in \cite{Bandres:2008ry}, we replace the Poincare supersymmetry parameter $\epsilon$ with $\gamma\cdot x \eta$ and add an additional term to the transformation of the spinor field
\begin{equation}
\begin{split}
\delta'\Psi_A &= \Omega_{AB} X^B \eta\\
\delta'\Psi^A &= \Omega^{AB} X_B \eta\;.
\end{split}
\end{equation}
Then it is straightforward to check that the Lagrangian is invariant under this superconformal symmetry.

Finally, let us briefly remark on the possibility of having $\mathcal{N}=5$ supersymmetry. That is, the supersymmetry generators transform as $\bf 5$ under $SO(5) \cong Sp(2)$. In this case, we require the Lagrangian be invariant under the supersymmetry transformations
\begin{equation}
\begin{split}
\delta X_A &= i (\Gamma^I_{AB} - \Omega_{AB}) \,\bar{\epsilon}^I \,\Psi^B\\
\delta \Psi_A &= (\Gamma^I_{AB} - \Omega_{AB}) \gamma^{\mu}\, \epsilon^I\, D_{\mu} X^B + \delta_3 \Psi_A\;,
\end{split}
\end{equation}
with their adjoints. We can follow the same procedure as above. The relations \eqref{E:CoefRel} follow as before since they do not involve the terms of the form $\Omega \Omega \psi \psi X X$. But when we next consider the variation due to the gauge boson $A_{\mu}$ in the $X$ field kinetic term and the spinor field $\Psi$ in $\cL_{4a,b,c}$ and $\cL'$, in addition to the terms
\begin{equation}
\begin{split}
2i (\tilde{\Gamma}^{IBC} &+ \Omega^{BC})(\bPsi_A \gamma^{\mu} \, \epsilon^I \, \alpha_1 D_{\mu} ( X_B X^A X_C) \\
&+i(\tilde{\Gamma}^{IBC} + \Omega^{BC})(\bPsi_B \gamma^{\mu} \, \epsilon^I \, (2\alpha_1-1) D_{\mu} (X_C X^A X_A - X_A X^A X_C)
\end{split}
\end{equation}
which can be canceled out by defining $\delta_3 \Psi^A$ just as before, we are left with additional terms
\begin{equation}
\begin{split}
-2 i(\alpha_1 -1) \tilde{\Gamma}^{IAB} \Omega^{CD} \bPsi_C \gamma^{\mu} \epsilon^I ( &\Omega_{AD} X_B X_E D_{\mu} X_E - \Omega_{AD} D_{\mu} X_E X^E X_B \\
&+\Omega_{AE} D_{\mu} X_B X^E X_D - \Omega_{AE} X_D X^E D_{\mu} X_B )
\end{split}
\end{equation}
which cannot be written in a form $(\tilde{\Gamma}^{IBC} + \Omega^{BC} ) \bPsi_A \gamma_{\mu} \, \epsilon^I \, D_{\mu} M_{BC}^A$ where $M_{BC}^A$ is a product of X.
This cannot be absorbed by a redefinition of $\delta_3 \bPsi^A$.
Therefore $\alpha_1=1$ and terms in \eqref{E:sp2inv} have to vanish. Then we get back to the $\mathcal{N}=6$ supersymmetric case. Therefore we conclude that the $\mathcal{N}=5$ supersymmetric Lagrangian whose supercharges are in the $\bf 5$ representation of $SO(5)$ does not exist.


%
%

\end{document}